\documentclass[aps,prd,twocolumn,superscriptaddress,nofootinbib]{revtex4-2}

\usepackage[margin=0.8in]{geometry}
\usepackage{amsmath,amssymb}
\usepackage{graphicx}
\usepackage{booktabs}
\usepackage{hyperref}
\usepackage{xcolor}

\usepackage{orcidlink}

\newcommand{\PU}{\affiliation{\small Phenikaa Institute for Advanced Study, Phenikaa University, Nguyen Trac, Duong Noi, Hanoi 100000, Vietnam}}

\newcommand{\NCTS}{\affiliation{\small Physics Division, National Center for Theoretical Sciences, National Taiwan University, Taipei 106319, Taiwan}}

\newcommand{\NTNU}{\affiliation{\small Department of Physics, National Taiwan Normal University, Taipei 116, Taiwan}}

\newcommand{\SC}{\affiliation{\small College of Physics, Sichuan University, Chengdu 610065, China}}

\begin{document}

\title{Probing the Axion--Photon--Dark Photon Interaction with LHC
Mono-Photons and Fermi-LAT Gamma-Ray Lines}

\author{Chuan-Ren Chen \orcidlink{0000-0002-3220-2550}} 
\email{crchen@ntnu.edu.tw} \NTNU

\author{Takaaki Nomura \orcidlink{0000-0002-0864-8333}}
\email{nomura@scu.edu.cn} \SC

\author{Van Que Tran \orcidlink{0000-0003-4643-4050}}
\email{vqtran@phys.ncts.ntu.edu.tw} \NCTS \PU

\date{\today}

\begin{abstract}
We investigate the axion--photon--dark photon interaction at the LHC and in gamma-ray line searches. 
By recasting the CMS mono-photon data, we derive a $95\%$ CL exclusion 
reaching $g_{a\gamma'\gamma}\simeq4.8\times10^{-4}\ {\rm GeV}^{-1}$ for dark photon masses $m_{\gamma'}\lesssim100$ GeV, with sensitivity extending to $m_{\gamma'}\simeq2.3$ TeV, where the effective theory approaches its validity limit. 
This improves upon the existing LEP mono-photon constraint by a factor of $3$ or more, depending on the dark photon mass. 
The projected HL-LHC sensitivity improves by a further factor of $\sim2.7$ and extends the reach to $m_{\gamma'}\simeq3$ TeV. 
The same interaction also induces $s$-wave annihilation of axions into $\gamma\gamma$. If the axion constitutes dark matter, Fermi-LAT gamma-ray line data become more constraining than the LHC for $m_a\gtrsim160$ GeV at $m_{\gamma'}=300$ GeV, where the on-shell collider signal is kinematically suppressed. Combining collider and gamma-ray line constraints excludes a thermal relic axion across the axion mass range considered for this benchmark.

\end{abstract}

\maketitle

\section{Introduction}
\label{sec:intro}

Axions ($a$), originally proposed to solve the strong-$CP$ problem, and dark
photons ($\gamma'$), the gauge bosons of hidden $U(1)$ symmetries appearing widely
in extensions of the Standard Model (SM), are both well-motivated,
feebly coupled additions to the SM particle content. Beyond the
familiar axion--photon--photon coupling and photon--dark photon kinetic
mixing, ultraviolet completions containing heavy fermions charged
under both the SM and the hidden sector can generate a three-way
vertex linking an axion, a photon, and a dark photon (with an effective coupling $g_{a\gamma'\gamma}$)~\cite{Kaneta:2016wvf}. 
This coupling was studied in detail at lepton colliders 
in Refs.~\cite{Biswas:2019lcp, Chen:2025hbh, Yue:2026oqu}. 
In particular, Ref.~\cite{Chen:2025hbh} showed that the processes $e^+e^-\to\gamma'\to\gamma a$ and
$e^+e^-\to\gamma^*/Z\to a\gamma'$ with $\gamma'\to a\gamma$, with the
axion escaping the detector, give LEP~II data and future facilities
(ILC, FCC-ee, CEPC) sensitivity to $g_{a\gamma'\gamma}$ down to
$\mathcal O(10^{-4})$~GeV$^{-1}$ for dark photon masses below the
collider energy.

The existing LHC mono-photon searches provide a complementary probe of this interaction in
the heavier mass region beyond the reach of lepton colliders. The present paper
carries out that analysis. At the LHC, the dominant
production channel is $q\bar q\to\gamma^*/Z^*\to a\gamma'$ through the
same portal vertex, followed by $\gamma'\to a\gamma$. 
If the axion is stable on detector timescales, as expected when it constitutes dark matter, 
it will escape the detector and give rise to a genuine mono-photon plus missing transverse momentum ($E_T^{\rm miss}$) signature, 
which is directly targeted by the CMS search.
We recast CMS's search for events with
a
single photon and large $E_T^{\rm miss}$ 
at
$\sqrt s=13$~TeV~\cite{CMS:2025cgw}.

We construct a full simulation chain using MadGraph5, Pythia8 and Delphes
 for $pp\to aa\gamma$, and
a CMS signal-region (SR) selection. We validate the chain against
CMS's own published benchmark efficiency, and construct a multi-bin
$CL_s$ statistical method from CMS's public per-bin data. 
Using this chain we derive a 95\% confidence level (CL) exclusion on
$g_{a\gamma'\gamma}$ as a function of  
the dark photon mass, compare it
with existing and projected lepton collider reach, and project it to
the High-Luminosity LHC (HL-LHC). Finally, if the axion itself is the
dark matter (DM), the same vertex opens $s$-wave annihilation into
monochromatic photons. We derive annihilation cross
sections and confront the collider limit with the Fermi-LAT gamma-ray
line search and with the thermal relic target.

The paper is organized as follows. Section~\ref{sec:model} summarizes
the model. Section~\ref{sec:method} describes the simulation chain,
its validation, and the statistical method. Section~\ref{sec:results}
presents the LHC results, including the HL-LHC projection and the
comparison with lepton colliders. Section~\ref{sec:dm} derives the
DM constraints, and Section~\ref{sec:conclusion} discusses
our results and concludes.

\section{The Model}
\label{sec:model}

We adopt the effective Lagrangian of Refs.~\cite{Kaneta:2016wvf, Chen:2025hbh},
\begin{align}
\mathcal L \supset{}& -\frac14 F_{\mu\nu}F^{\mu\nu}
- \frac14 F'_{\mu\nu}F'^{\mu\nu} \notag\\
&- \frac{\epsilon}{2c_W}F_{\mu\nu}F'^{\mu\nu}
+ \frac{g_{a\gamma'\gamma}}{2c_W}\,a\,F_{\mu\nu}\tilde F'^{\mu\nu},
\label{eq:lagrangian}
\end{align}
where $F$ and $F'$ are the hypercharge and dark photon field-strength
tensors, $\tilde F'^{\mu\nu}=\frac12\epsilon^{\mu\nu\rho\sigma}
F'_{\rho\sigma}$, $c_W$ ($s_W$, $t_W$) is the cosine (sine, tangent) of
the weak mixing angle, $\epsilon$ is the kinetic-mixing parameter, and
$a$ is a pseudoscalar axion of mass $m_a$. 
Here we consider the dark photon of mass $m_{\gamma'}$ where 
its mass can be generated either by a Stueckelberg
mechanism or by the spontaneous breaking of the hidden $U(1)'$ through
the vacuum expectation value of an SM singlet dark Higgs
field~\cite{Fabbrichesi:2020wbt}
\footnote{We take the dark Higgs, if present,
to be heavy enough that it plays no role in the phenomenology
discussed here.}. 
The dimension-five operator
with coefficient $g_{a\gamma'\gamma}$ is the interaction of interest.
Because $F$ is the hypercharge field strength, after electroweak
symmetry breaking it contains both an $a\gamma'\gamma$ vertex, with
coupling $g_{a\gamma'\gamma}$, and an $a\gamma'Z$ vertex, with coupling
of magnitude $t_W\,g_{a\gamma'\gamma}$. It gives the dark photon
decay channels~\cite{Chen:2025hbh},
\begin{align}
\Gamma(\gamma'\to a\gamma) &= \frac{g_{a\gamma'\gamma}^2}{96\pi}
  m_{\gamma'}^3\left(1-\frac{m_a^2}{m_{\gamma'}^2}\right)^3,
  \label{eq:gammaagam}\\
\Gamma(\gamma'\to aZ) &= \frac{t_W^2g_{a\gamma'\gamma}^2}{96\pi}
  m_{\gamma'}^3\Big\{\Big[1-\Big(\tfrac{m_Z-m_a}{m_{\gamma'}}\Big)^2\Big]
  \notag\\
  &\quad\times\Big[1-\Big(\tfrac{m_Z+m_a}{m_{\gamma'}}\Big)^2\Big]
  \Big\}^{3/2}.
  \label{eq:gammaaz}
\end{align}
The $aZ$ channel is suppressed by $t_W^2$ and by phase space. 
The two channels dominate over the kinetic-mixing-induced 
decays into SM fermions when
$m_{\gamma'}g_{a\gamma'\gamma} \gg \epsilon$, making $\gamma'\to
a\gamma$ the leading decay mode. We work throughout in this regime,
consistent with the assumption adopted in Ref.~\cite{Chen:2025hbh} for its constraints 
in the
$(m_{\gamma'},g_{a\gamma'\gamma})$ plane. The kinetic-mixing-induced
production $q\bar q\to\gamma'^{(*)}\to a\gamma$ ($\propto\epsilon^2$)
is then also negligible.

At the LHC, the axion and dark photon can be produced 
via $q\bar q\to\gamma^*/Z^{(*)}\to a\gamma'$. 
The dark photon can then decay to an axion and a photon. 
The result is a
resonant topology with two invisible axions and one visible, on-shell
photon: the exact final state targeted by CMS's search~\cite{CMS:2025cgw}.
This topology differs qualitatively from the simplified-DM and
contact-operator benchmarks CMS uses for its own signal
interpretation. Their
quoted efficiencies therefore cannot be assumed to transfer to this
model. We validate our own chain independently in
Sec.~\ref{sec:method-validation}. Unless stated otherwise we fix
$m_a=1$~GeV, a representative light axion benchmark. In
Sec.~\ref{sec:dm} we scan $m_a$ at fixed $m_{\gamma'}=300$~GeV.

\section{Simulation and Analysis Method}
\label{sec:method}

\subsection{Signal generation}
\label{sec:method-gen}

We implement the Lagrangian of Eq.~\eqref{eq:lagrangian} as a UFO
model and generate the full $2\to3$ process $pp\to a\,a\,\gamma$ at
leading order (LO) with MadGraph5~\cite{Alwall:2014hca}. 
Since we assume that the
axion couples only through the vertex of Eq.~\eqref{eq:lagrangian},
and the kinetic mixing is set to a negligible value, this
final state arises only from $q\bar q\to\gamma^*/Z^*\to a\gamma'$
followed by $\gamma' \to a\gamma$. The dark photon is kept as a
propagator with its total width recomputed at every parameter point.

We do not factorize production
and decay using the narrow-width approximation. 
This is because, as follows from
Eqs.~\eqref{eq:gammaagam} and~\eqref{eq:gammaaz}, 
the dark photon width grows with its mass and the coupling,
$\Gamma_{\gamma'}\propto g_{a\gamma'\gamma}^2m_{\gamma'}^3$, and the resonance is no longer narrow at the couplings probed
for sufficiently large dark photon masses. The off-shell dark photon contributions
are then included automatically.

Following CMS's own signal simulation~\cite{CMS:2025cgw}, we apply a
generator-level photon pre-selection on transverse energy and rapidity, $E_T^\gamma>130$~GeV and
$|\eta^\gamma|<2.5$. It removes only events that cannot pass the SR
photon requirement described below. It also concentrates the simulated events in
the SR relevant phase space, including at the low dark photon masses,
where most of the cross section lies at low photon $E_T$. We generate $4\times10^4$ events per
parameter point, shower them with Pythia8~\cite{Sjostrand:2014zea},
and pass them through Delphes~\cite{deFavereau:2013fsa} with its CMS detector card.

Our SR selection is implemented on the Delphes output
following CMS's published cuts~\cite{CMS:2025cgw}:
\begin{itemize}
\item one photon with $E_T^\gamma>225$~GeV and $|\eta^\gamma|<1.44$,
  passing the $H/E$ and isolation requirements;
\item $E_T^{\rm miss}>200$~GeV;
\item $\Delta\phi>0.5$ between $E_T^{\rm miss}$ and both the photon
  and each of the four leading jets;
\item $E_T^\gamma/E_T^{\rm miss}<1.4$;
\item a veto on electrons and muons.
\end{itemize}
Here $H/E$ is the ratio of the hadronic calorimeter energy behind
the photon to its ECAL energy ($H/E < 0.023$), and the isolation requirements bound the
summed transverse momenta of charged hadrons, neutral hadrons, and
other photons in a cone of $\Delta R<0.3$ around the photon. CMS
additionally requires a narrow shower in the ECAL,
$\sigma_{i\eta i\eta}<0.0103$~\cite{CMS:2025cgw}, 
where $\sigma_{i\eta i\eta}$ is the energy-weighted width of the shower in
the pseudorapidity ($\eta$) computed over the $5\times5$ crystals around the most energetic
crystal~\cite{CMS:2015myp}. This variable separates prompt photons
from the wider showers of hadronic activity (e.g. $\pi^0\to\gamma\gamma$).
Delphes does not model ECAL shower shapes, so a flat $80\%$ photon
identification scale factor is applied instead of the
$\sigma_{i\eta i\eta}$ requirement, consistent with our
validation benchmark shown in the next section.

\subsection{Validation against CMS's benchmarks}
\label{sec:method-validation}

Before applying the simulation chain to our signal, 
we validate it against two benchmark scenarios considered in the CMS analysis~\cite{CMS:2025cgw}. 
First, we consider the electroweak-DM (EW-DM) contact interaction benchmark, 
$q\bar q\to Z/\gamma^*\to\gamma\chi\bar\chi$, mediated by a dimension-seven operator~\cite{Nelson:2013pqa}, 
at the $m_\chi=1$~GeV and $m_\chi=400$~GeV benchmark points. 
HEPData reports overall efficiencies of $36.45\%$ and $42.01\%$, respectively~\cite{CMS:2025cgw}. 
Implementing the same operator in a UFO model and passing it through our simulation chain, 
we obtain efficiencies of $39.5\%$ and $43.3\%$, corresponding to differences of approximately 
$8\%$ and $3\%$, respectively. We further validate our framework against 
the simplified-DM benchmark considered in the same CMS analysis, 
for which our overall efficiency agrees with the CMS result to within $4\%$. 
The residual differences are attributable to the parameterized detector response 
and fast simulation approximations inherent to Delphes~\cite{deFavereau:2013fsa}. 
A detailed validation is provided in Appendix~\ref{app:validation}.

\subsection{Statistical method}
\label{sec:method-stats}

We set 95\% CL limits using the $CL_s$
method~\cite{Junk:1999kv, Read:2002hq}. 
CMS's HEPData record~\cite{CMS:2025cgw} does not
provide a full likelihood or background covariance. We therefore construct 
the statistical treatment using the most granular publicly available information:
SR yields in five $E_T^\gamma$ bins ($[225,250]$, $[250,300]$,
$[300,400]$, $[400,600]$, and $[600,1000]$~GeV) in each of CMS's two
SR sub-regions, which are split by photon azimuthal angle $\phi$ to isolate a
beam-halo background. The region with $|\phi| < 0.5$ is called the horizontal region, 
and the rest is called the vertical region. 
This gives ten independent Poisson counting
channels, each with its own post-fit background yield and uncertainty.
We combine these channels into a likelihood-ratio test statistic and
derive an upper limit on the signal strength using toy Monte Carlo. 
Each bin's background is Gaussian-constrained around its quoted central value; no
cross-bin correlations are included.

The fit yields an upper limit $\sigma_{95}$ (the cross section for which $CL_s=0.05$) on the pre-selected
signal cross section, given the SR efficiency and the $E_T^\gamma$
shape of the simulated sample.  Cross sections above
$\sigma_{95}$ are excluded at $95\%$ CL.
In the narrow-width limit the cross section scales as
$g_{a\gamma'\gamma}^2$, since the $g_{a\gamma'\gamma}^4$ of the
amplitude squared is compensated by the $1/\Gamma_{\gamma'}\propto
g_{a\gamma'\gamma}^{-2}$ of the resonance. Once $\Gamma_{\gamma'}$
is comparable to $m_{\gamma'}$ this is no longer true, and the cross
section eventually saturates. We therefore compute the pre-selected LO cross section
$\sigma(g_{a\gamma'\gamma})$ on a grid of fourteen couplings between
$10^{-4}$ and $10^{-1}$~GeV$^{-1}$ at each mass point, and define the
limit by
\begin{equation}
K\,\sigma\big(g_{a\gamma'\gamma}^{(95)}\big)=\sigma_{95},
\label{eq:rescale}
\end{equation}
interpolating $\ln\sigma$ linearly in $\ln g_{a\gamma'\gamma}$. If
$K\sigma$ stays below $\sigma_{95}$ for every coupling, the mass point
is not excluded. The efficiency and shape are taken from the
$4\times10^4$-event sample, generated at a coupling close to the
resulting limit, so the width in the sample matches the width at the
limit. Here $K$ is the NLO normalization factor. We use the commonly
quoted value for Drell-Yan-like $q\bar q$-initiated production at LHC
energies, $K=1.3$, as a single overall rescaling of the LO cross
section.

\begin{figure}[t]
\centering
\includegraphics[width=\columnwidth]{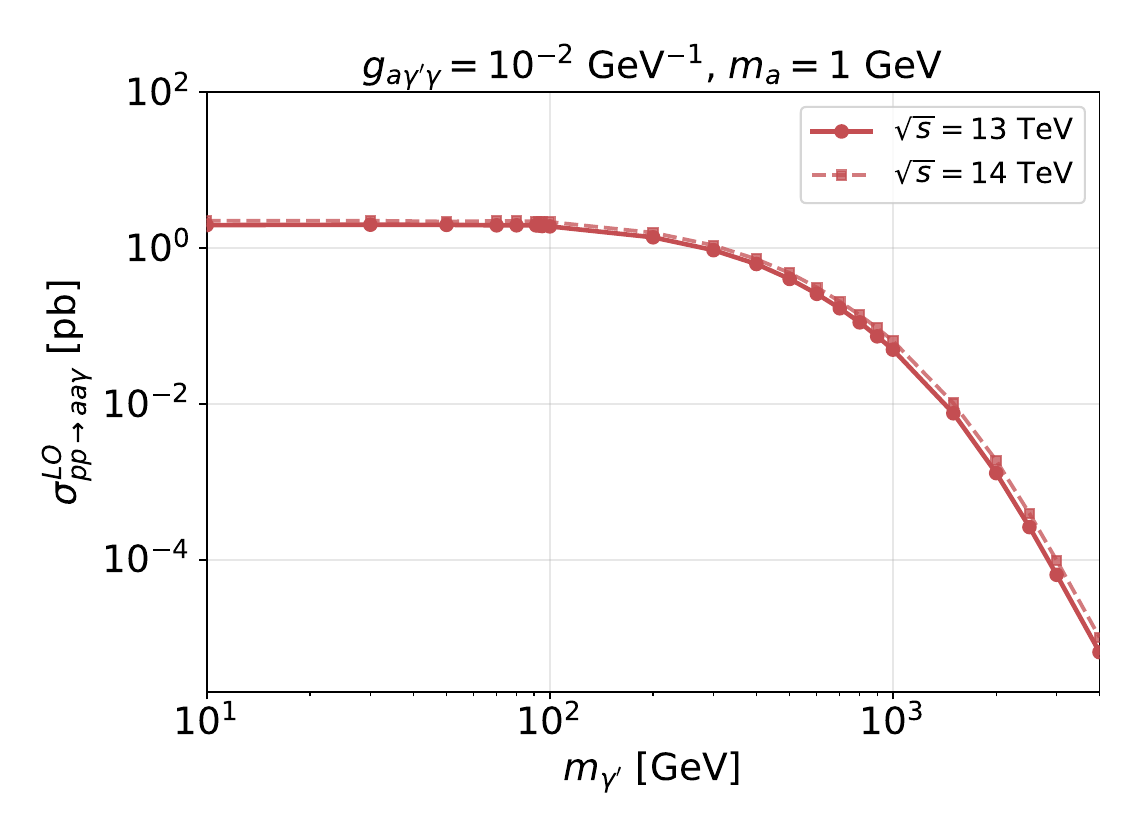}
\caption{LO signal cross section $\sigma(pp\to aa\gamma)$ after the
generator-level photon pre-selection ($E_T^\gamma>130$~GeV,
$|\eta^\gamma|<2.5$), as a
function of dark photon mass at $g_{a\gamma'\gamma}=10^{-2}$~GeV$^{-1}$
and $m_a=1$~GeV, for $\sqrt s=13$~TeV (solid) and $14$~TeV (dashed).}
\label{fig:inclusivexsec}
\end{figure}

\begin{figure*}[tb]
\centering
\includegraphics[width=\textwidth]{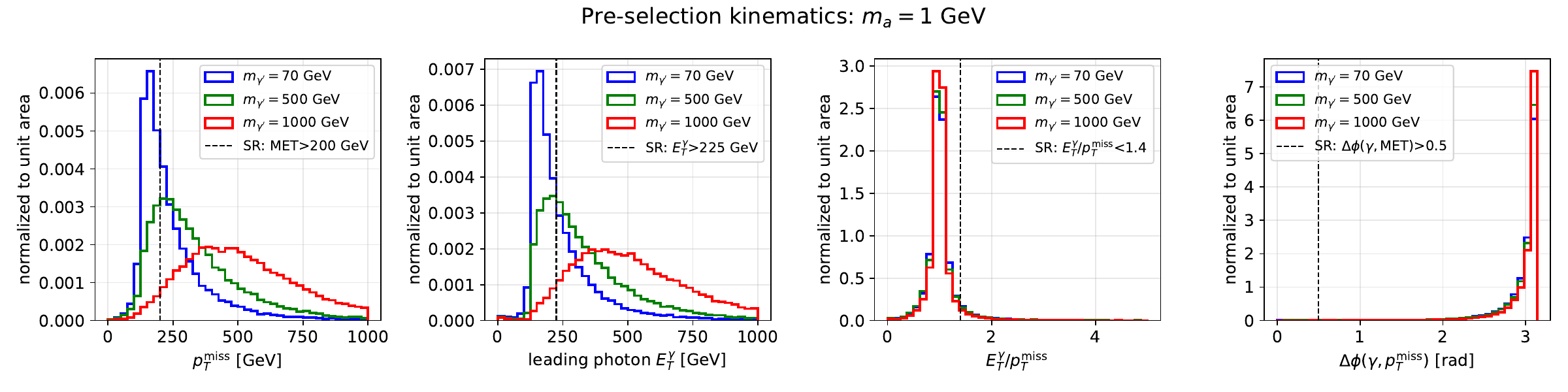}
\caption{Kinematics of the leading photon
candidate (no identification or isolation applied) in the
pre-selected samples ($E_T^\gamma>130$~GeV, $|\eta^\gamma|<2.5$ at
generator level) for
$m_{\gamma'}=70,500,1000$~GeV and $m_a=1$~GeV, normalized to unit
area. Dashed lines mark the corresponding SR thresholds.}
\label{fig:kinmasscomp}
\end{figure*}

\section{LHC Results}
\label{sec:results}

\subsection{Signal cross section and pre-selection kinematics}
\label{sec:results-xseckin}

Two ingredients shape the exclusion limit: how the production cross
section falls with $m_{\gamma'}$, and how the event kinematics harden
with it. Fig.~\ref{fig:inclusivexsec} shows the signal cross
section after the photon pre-selection across the scanned range of
dark photon masses, at $g_{a\gamma'\gamma}=10^{-2}$~GeV$^{-1}$ and
$m_a=1$~GeV. Below $m_{\gamma'}\simeq100$~GeV it is nearly
independent of the dark photon mass, $\sigma\simeq2.0$~pb at
$\sqrt{s}=
13$~TeV. For $m_{\gamma'}+m_a<m_Z$ the intermediate $Z$ in
$q\bar q\to Z^{(*)}/\gamma^*\to a\gamma'$ can go on shell and the inclusive
rate is resonantly enhanced. However, the photon from that
configuration carries at most $E_\gamma\simeq(m_Z^2-m_{\gamma'}^2)/(2m_Z)
<46$~GeV in the $Z$ rest frame and is removed by the $E_T^\gamma>130$~GeV
requirement. The pre-selected rate instead comes from far off-shell
$\gamma^*/Z^*$ at large partonic energy, which is insensitive to a light
$m_{\gamma'}$. Above $m_{\gamma'}\simeq100$~GeV the cross section
falls steadily, to about $0.05$~pb at $1$~TeV and $1.3\times10^{-3}$~pb at
$2$~TeV.
Going to $\sqrt s=14$~TeV raises the cross section by a factor of
$1.13$ at light dark photon masses and $1.30$ at $1$~TeV.

The other half of the story is how efficiently the SR selection retains
events at each mass. Fig.~\ref{fig:kinmasscomp} compares the
kinematics of the leading photon candidate in the pre-selected
samples (before any SR requirement)
at three representative masses, $m_{\gamma'}=70,500,1000$~GeV, each
normalized to unit area. For $m_{\gamma'}=70$~GeV, the
distributions are sculpted by the
pre-selection. The missing transverse momentum and the leading-photon
$E_T$ peak at just above the $130$~GeV threshold, with medians of
about $200$~GeV. The distributions harden for heavier dark photons,
with medians of about $290$~GeV  and
$500$~GeV for $m_{\gamma'}=500$~GeV and  $m_{\gamma'}=1000$~GeV, respectively, as a heavier resonance
splits its energy into a harder visible photon and a harder invisible
recoil.
By contrast, the $E_T^\gamma/E_T^{\rm miss}$ and
$\Delta\phi(\gamma,E_T^{\rm miss})$ distributions are nearly
mass-independent, peaking at $\simeq1$ and $\pi$, respectively, since
both reflect the same back-to-back topology at every energy scale.

\begin{figure*}[t]
\centering
\includegraphics[width=0.95\textwidth]{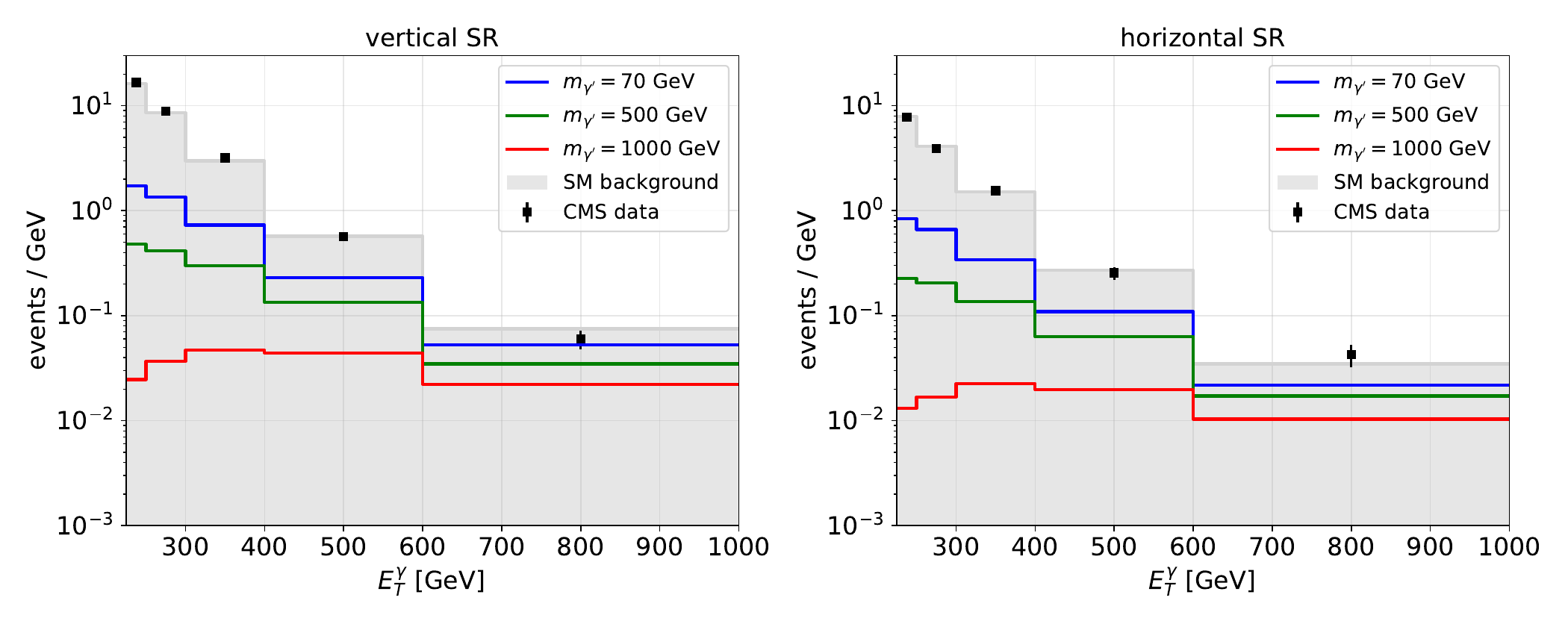}
\caption{Predicted signal $E_T^\gamma$ spectra, in events per GeV, for $m_{\gamma'}=70,500,1000$~GeV at
$g_{a\gamma'\gamma}=10^{-3}$~GeV$^{-1}$ and $m_a=1$~GeV, overlaid on
CMS's post-fit background (gray) and observed data (black points) in the
vertical (left) and horizontal (right) SR sub-regions, for
$101$~fb$^{-1}$~\cite{CMS:2025cgw}.}
\label{fig:sroverlay}
\end{figure*}

\subsection{Signal in the CMS signal region}
\label{sec:results-sr}

Table~\ref{tab:cutflow} gives the cumulative SR cutflow at the same
three representative masses, relative to the pre-selected sample. The photon
requirement $E_T^\gamma>225$~GeV, $|\eta^\gamma|<1.44$ is by far the
dominant loss and is responsible for the rise of the efficiency with
$m_{\gamma'}$, from $25.5\%$ at $70$~GeV to $73.1\%$ at $1000$~GeV.
The subsequent cuts remove comparable fractions of the surviving events
at all three masses: $14$--$15\%$ (relative) for isolation, $9$--$12\%$
for the jet--$E_T^{\rm miss}$ $\Delta\phi$ requirement, and
$\simeq2\%$ for the $E_T^\gamma/E_T^{\rm miss}$ ratio. The
$E_T^{\rm miss}>200$~GeV cut is the only one with a notable mass
dependence, removing $9\%$ of the remaining events at
$m_{\gamma'}=70$~GeV but only $2$--$5\%$ at
$500$--$1000$~GeV. The other cuts remove $\lesssim0.1\%$. After applying the
photon identification scale factor, the final efficiency rises from $14.3\%$ at
$m_{\gamma'}=70$~GeV to $42.7\%$ at $1000$~GeV.

\begin{table}[t]
\centering
\caption{Cumulative SR cutflow efficiency (\%) at $m_a=1$~GeV,
relative to the generator-level pre-selected sample
($E_T^\gamma>130$~GeV, $|\eta^\gamma|<2.5$). The final row includes
the $80\%$ photon identification scale factor.}
\label{tab:cutflow}
\begin{tabular}{lrrr}
\toprule
 & \multicolumn{3}{c}{$m_{\gamma'}$ [GeV]} \\
Cut & 70 &  500 & 1000 \\
\midrule
$E_T^\gamma>225$~GeV, $|\eta^\gamma|<1.44$ & 25.5 &  49.2 & 73.1 \\
$+\,H/E<0.023$ & 25.5 &  49.2 & 73.1 \\
$+$ isolation & 22.0 &  42.0 & 63.0 \\
$+\,E_T^{\rm miss}>200$~GeV & 20.0 &  40.0 & 61.9 \\
$+\,\Delta\phi(E_T^{\rm miss},\gamma)>0.5$ & 20.0 &  40.0 & 61.9 \\
$+\,\Delta\phi(E_T^{\rm miss},j)>0.5$ & 18.2 &  36.0 & 54.5 \\
$+\,E_T^\gamma/E_T^{\rm miss}<1.4$ & 17.8 &  35.3 & 53.4 \\
$+$ lepton veto & 17.8 &  35.3 & 53.4 \\
$\times$ photon ID scale factor & \textbf{14.3} &  \textbf{28.2} & \textbf{42.7} \\
\bottomrule
\end{tabular}
\end{table}

Fig.~\ref{fig:sroverlay} overlays the resulting signal $E_T^\gamma$
spectra on CMS's post-fit background and observed data in both SR
sub-regions~\cite{CMS:2025cgw}. The signal yields are computed 
for $g_{a\gamma'\gamma}=10^{-3}$~GeV$^{-1}$, $m_a=1$~GeV, 
and the luminosity of $101$~fb$^{-1}$.  The total predicted signal is
$370$,  $152$, and $37$ events for $m_{\gamma'}=70$, 
$500$, and $1000$~GeV, respectively. For comparison, the total
background is $1897$ events ($1275$ vertical $+$ $622$ horizontal)
and $1928$ events are observed~\cite{CMS:2025cgw}.

Fig.~\ref{fig:sroverlay} shows where the sensitivity comes from at
different masses. At every mass the signal falls more slowly with
$E_T^\gamma$ than the background, whose rate per GeV drops by a factor
of $\sim200$ between the first and last bins. The signal-to-background
ratio therefore rises with $E_T^\gamma$, and the $[400,600]$ and
$[600,1000]$~GeV bins carry most of the sensitivity.  
For a light dark photon the larger production cross section makes up for the
lower efficiency. Only the boosted tail passes the
$E_T^\gamma>225$~GeV and $E_T^{\rm miss}>200$~GeV requirements, and
about $26\%$ of the surviving signal lies in the two highest
bins. For a heavy dark photon the harder spectrum raises this fraction to
$40\%$ at $m_{\gamma'}=500$~GeV and $71\%$ at $1000$~GeV. 
This motivates the use of the multi-bin likelihood 
rather than a single-bin counting analysis,
especially for heavy dark photon masses.

\subsection{Dark photon mass scan}
\label{sec:results-map}

\begin{figure}[t]
\centering
\includegraphics[width=\columnwidth]{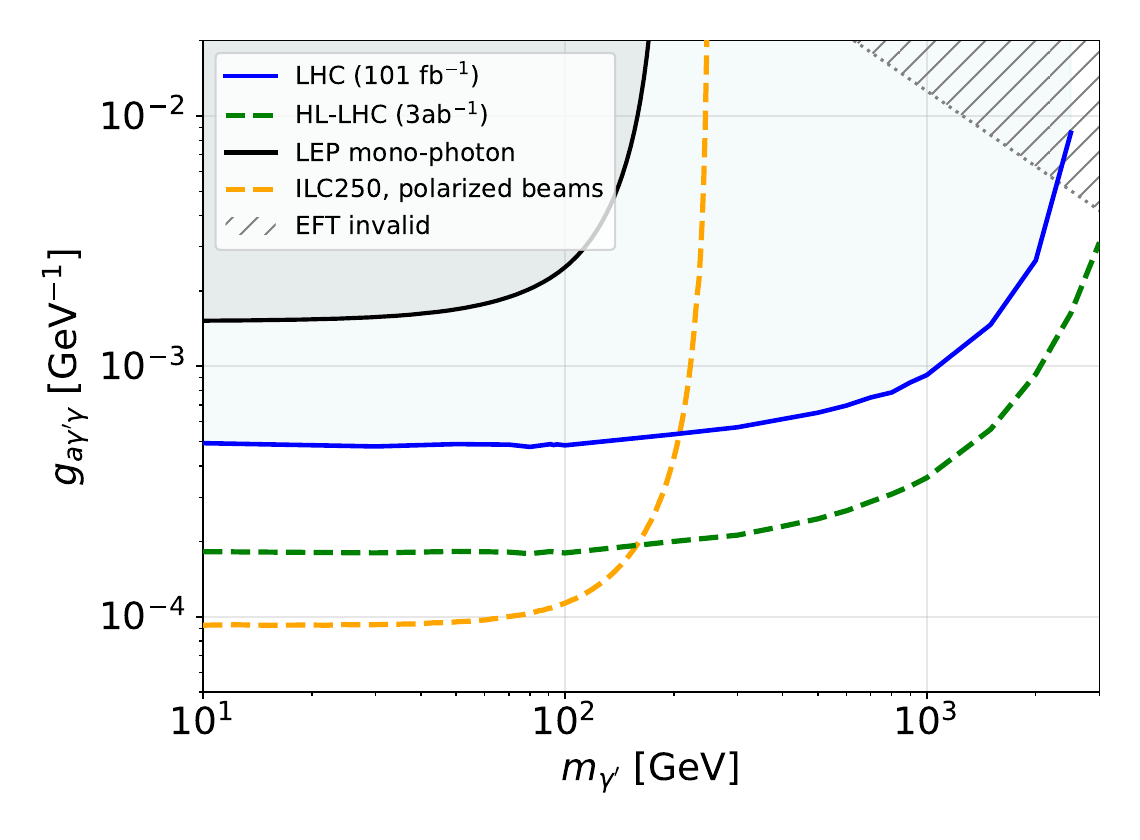}
\caption{95\% CL exclusion on $g_{a\gamma'\gamma}$ vs.\ $m_{\gamma'}$
at $m_a=1$~GeV: this work with current LHC data (solid blue, shaded
region excluded) and projected to the HL-LHC (dashed green). The hatched region,
$g_{a\gamma'\gamma}m_{\gamma'}>4\pi$, is where the effective theory is
no longer perturbative. Also shown are the LEP mono-photon constraint (solid
black) and the projected sensitivity of a $250$~GeV ILC with polarized
beams (dashed orange), both from Ref.~\cite{Chen:2025hbh}.}
\label{fig:hllhc}
\end{figure}

We scan $m_{\gamma'}=10$--$4000$~GeV with $m_a=1$~GeV 
and convert each point to a 95\% CL
exclusion on $g_{a\gamma'\gamma}$ via Eq.~\eqref{eq:rescale}.
The
LHC exclusion curve is shown by the blue line in Fig.~\ref{fig:hllhc}. For
$m_{\gamma'}\lesssim100$~GeV both the pre-selected cross section and
the SR efficiency ($\simeq14\%$) are nearly mass independent, and the
limit is approximately flat at $g_{a\gamma'\gamma}^{(95)}\simeq4.8\times10^{-4}$
~GeV$^{-1}$, the strongest sensitivity in the scan. The limit then
weakens gradually, to $6.5\times10^{-4}$~GeV$^{-1}$ at $500$~GeV and
$9.2\times10^{-4}$~GeV$^{-1}$ at $1$~TeV, as the rising, then
plateauing, efficiency partly offsets the falling cross section. Up to
$1$~TeV the dark photon is narrow at the limit,
$\Gamma_{\gamma'}/m_{\gamma'}<0.4\%$, and the finite width plays no role.
It becomes important at higher masses. At $m_{\gamma'}=2$~TeV the limit,
$2.6\times10^{-3}$~GeV$^{-1}$, corresponds to
$\Gamma_{\gamma'}/m_{\gamma'}=12\%$. 
Once $\Gamma_{\gamma'}$ becomes comparable to
$m_{\gamma'}$, increasing the coupling enhances 
both the production amplitude and the total width, 
causing the cross section to saturate. 
For $m_{\gamma'}\geq3$~TeV the saturated rate stays below the excluded
one, reaching only $27\%$ ($3\%$) of $\sigma_{95}$ at $3$ ($4$)~TeV,
so no coupling is excluded there. 

The large couplings at the high mass end of the scan also test the
validity of the effective description itself. The interaction of
Eq.~\eqref{eq:lagrangian} is a dimension-five operator. The coupling
is generated by heavy states of mass $\Lambda$, with
$g_{a\gamma'\gamma}\sim c/\Lambda$ and $c\lesssim4\pi$ for a perturbative
ultraviolet completion. The effective operator is valid only for processes 
with characteristic energies below $\Lambda$, so
$g_{a\gamma'\gamma}\sqrt{\hat s}\lesssim4\pi$, and at least
$g_{a\gamma'\gamma}m_{\gamma'}\lesssim4\pi$, since
$\sqrt{\hat s}\geq m_{\gamma'}$. Beyond this boundary, 
the heavy states may become resolved or directly produced, 
and the rates predicted by
Eq.~\eqref{eq:lagrangian} no longer apply. The region
$g_{a\gamma'\gamma}m_{\gamma'}>4\pi$ is shown hatched in
Fig.~\ref{fig:hllhc}. The LHC exclusion lies inside the valid region up
to $m_{\gamma'}\simeq2.3$~TeV.

\paragraph{HL-LHC projection.} We project the exclusion to an integrated luminosity of
$3000$~fb$^{-1}$ at $\sqrt s=14$~TeV. The signal normalization uses the
pre-selected cross section $\sigma(g_{a\gamma'\gamma})$ computed at
$14$~TeV on the same coupling grid, while the SR efficiency and
$E_T^\gamma$ shape are taken from the $13$~TeV simulation. At
$m_{\gamma'}>2$~TeV, where the width at the HL-LHC limit differs
appreciably from that at the current one, these $13$~TeV samples are
generated at the HL-LHC coupling. 
CMS's per-bin
background, which corresponds to $101$~fb$^{-1}$, is scaled up by the luminosity ratio $3000/101$, and an Asimov
(background-only) dataset gives the median expected limit. Because the
background uncertainties at $3000$~fb$^{-1}$ are not known a priori, 
we assume today's per-bin uncertainty is
purely statistical and shrinks as $1/\sqrt{\mathcal L}$. 
The improvement is
nearly uniform up to $1.5$~TeV with a factor of $2.6$--$2.8$ on
$g_{a\gamma'\gamma}$. This combines the $\mathcal L^{1/4}$ scaling
expected for a background-dominated counting analysis,
$(3000/101)^{1/4}\simeq2.3$, with the $13$--$30\%$ larger cross
section at $14$~TeV.
The HL-LHC excludes $g_{a\gamma'\gamma}>3.1\times10^{-3}$~GeV$^{-1}$ at $m_{\gamma'}=3$~TeV
($g_{a\gamma'\gamma}m_{\gamma'}\simeq9<4\pi$), and its limit reaches the
validity boundary of the effective theory at $m_{\gamma'}\simeq3.1$~TeV. 
The HL-LHC sensitivity curve is shown by the dashed green line in Fig.~\ref{fig:hllhc}.

\paragraph{Comparison with lepton colliders.} Fig.~\ref{fig:hllhc}
also shows two lepton-collider curves from Ref.~\cite{Chen:2025hbh}.
The first is the existing LEP~II mono-photon constraint (solid black), which
covers $m_{\gamma'}\lesssim180$~GeV. The second is the projected
sensitivity of a $\sqrt s=250$~GeV ILC with polarized beams (dashed orange),
which extends to that machine's kinematic limit,
$m_{\gamma'}\simeq 250$~GeV. Our LHC limit is stronger than the LEP
bound across the entire range that bound covers. 
The improvement factor is approximately three at
 $m_{\gamma'}=10$~GeV, grows to an order of magnitude by
$m_{\gamma'}\simeq140$~GeV, and exceeds $100$ as the LEP curve
approaches its kinematic endpoint.

By contrast, ILC250 would provide greater sensitivity for light
dark photon. It is about $5$ times stronger than today's LHC limit
at $m_{\gamma'}=10$~GeV, remains stronger than it up to
$m_{\gamma'}\simeq 205$~GeV, and stronger than the HL-LHC projection up
to $m_{\gamma'}\simeq157$~GeV. Its sensitivity collapses as
$m_{\gamma'}$ approaches $\sqrt s$. Above $m_{\gamma'}\simeq250$~GeV,
only a hadron collider can probe this interaction, and the present
recast is, to our knowledge, the first
constraint there. The two machine types are therefore complementary:
a future lepton collider has the reach at light $m_{\gamma'}$, and the
LHC and HL-LHC alone cover $m_{\gamma'}> 250$~GeV, up to
$m_{\gamma'}\simeq2.3$~TeV ($3$~TeV) for the LHC (HL-LHC), where the
effective theory reaches its validity limit.

\section{Dark Matter Constraints}
\label{sec:dm}

The dark axion is electrically neutral 
and can serve as a DM candidate in this model 
if it is stable on cosmological timescales. 
For nonzero $\epsilon$, however, the axion can decay 
via $a\to\gamma\gamma'^{*}\to\gamma f\bar f$. 
This decay is forbidden by a $Z_2$ symmetry 
under which $a\to-a$ and $\gamma'\to-\gamma'$, 
while all SM fields are even. 
The symmetry leaves the $a\gamma'\gamma$ operator 
in Eq.~\eqref{eq:lagrangian} invariant 
but forbids the kinetic-mixing term. 
Consequently, the lightest $Z_2$-odd state, 
which is the axion for $m_a<m_{\gamma'}$, 
is absolutely stable. In this case, $\epsilon=0$ exactly. 
Our collider results, however, 
apply more generally to any $\epsilon\ll m_{\gamma'}g_{a\gamma'\gamma}$, 
including the exactly stable case.

\subsection{Annihilation cross sections}
\label{sec:dm-xsec}

If the axion
makes up the observed DM, the same vertex opens the annihilation
channels $aa\to\gamma\gamma$, $\gamma Z$, and $ZZ$ through $t$- and
$u$-channel $\gamma'$ exchange. Each diagram contains the vertex twice,
so $\langle\sigma v\rangle\propto g_{a\gamma'\gamma}^4$. Annihilation
at rest into $\gamma\gamma$ produces monochromatic photons 
at $E_\gamma=m_a$, the gamma-ray line signature targeted
by DM indirect detection experiments~\cite{Fermi-LAT:2015kyq, HAWC:2019jvm, MAGIC:2022acl, CTAO:2024wvb, VERITAS:2024usn, HESS:2026ila}. 
The $aa\to\gamma'
\gamma'$ channel, through photon or $Z$ exchange, is closed for
$m_a<m_{\gamma'}$, which holds throughout this paper.

We derive the annihilation cross sections from the model's vertex
Feynman rules. Both the $a\gamma'\gamma$ and $a\gamma'Z$ vertices have
the $\epsilon$-tensor structure $\epsilon^{\mu\nu\rho\sigma}
k_{1\rho}k_{2\sigma}$ of Eq.~\eqref{eq:lagrangian}, with couplings of
magnitude $g_\gamma=g_{a\gamma'\gamma}$ and $g_Z=t_W\,g_{a\gamma'\gamma}$.
The longitudinal part of the $\gamma'$ propagator vanishes when
contracted into either vertex, since the same momentum then appears
twice in a totally antisymmetric tensor. At threshold, every Mandelstam
invariant is independent of the scattering angle, and the three
channels take a single closed form,
\begin{equation}
\langle\sigma v\rangle_{XY}=S_{XY}\,
\frac{g_X^2\,g_Y^2\;m_a\,p_{XY}^5}
{8\pi\left[m_a^2+m_{\gamma'}^2-\tfrac12(m_X^2+m_Y^2)\right]^2},
\label{eq:sigmav}
\end{equation}
where $XY=\gamma\gamma,\gamma Z,ZZ$ and $p_{XY}$ is the final-state
momentum in the center-of-mass frame. The factor $S_{XY}=1$ for
identical final-state bosons ($\gamma\gamma$, $ZZ$) and $S_{\gamma Z}=
2$, the latter reflecting the absence of the identical-particle factor
$\tfrac12$. Explicitly,
\begin{align}
\langle\sigma v\rangle_{\gamma\gamma}&=\frac{g_{a\gamma'\gamma}^4\,
  m_a^6}{8\pi\left(m_a^2+m_{\gamma'}^2\right)^2},
  \label{eq:sigmavgg}\\
\langle\sigma v\rangle_{\gamma Z}&=\frac{t_W^2\,g_{a\gamma'\gamma}^4\,
  m_a\,p_{\gamma Z}^5}{4\pi\left(m_a^2+m_{\gamma'}^2-m_Z^2/2\right)^2},
  \label{eq:sigmavgz}\\
\langle\sigma v\rangle_{ZZ}&=\frac{t_W^4\,g_{a\gamma'\gamma}^4\,
  m_a\,p_{ZZ}^5}{8\pi\left(m_a^2+m_{\gamma'}^2-m_Z^2\right)^2},
  \label{eq:sigmavzz}
\end{align}
with $p_{\gamma Z}=(4m_a^2-m_Z^2)/(4m_a)$ and $p_{ZZ}=\sqrt{m_a^2-m_Z^2}$.
The $\gamma Z$ ($ZZ$) channel opens for $m_a>m_Z/2$ ($m_a>m_Z$).
Because $\langle\sigma v\rangle$ is velocity-independent, the annihilation is $s$-wave,
as expected for this pseudoscalar $\epsilon$-tensor coupling. The same
cross section therefore controls freeze-out and present-day
annihilation in galactic halos, where $v\sim10^{-3}$. 

The $\gamma Z$ and $ZZ$
channels are subdominant but not negligible. At $m_{\gamma'}=300$~GeV
their ratios to $\gamma\gamma$ are $0.21$ and $10^{-3}$ at
$m_a=100$~GeV, rising to $0.56$ and $0.08$ at $m_a=290$~GeV. The
$\gamma Z$ photon line sits at $E_\gamma=p_\gamma<m_a$, spectrally
distinct from the $\gamma\gamma$ line. It is therefore not folded into
the line-search constraint below, but it contributes to the total
annihilation rate that sets the relic abundance.

\subsection{Thermal relic target}
\label{sec:dm-thermal}

If the axion was in thermal equilibrium in the early Universe and its
abundance was set by freeze-out through these channels, 
the observed DM density 
corresponds to the canonical $s$-wave value
$\langle\sigma v\rangle_{\rm tot}=\langle\sigma v\rangle_{\gamma\gamma}
+\langle\sigma v\rangle_{\gamma Z}+\langle\sigma v\rangle_{ZZ}\simeq
3\times10^{-26}$~cm$^3$\,s$^{-1}$. Because $\langle\sigma v\rangle_{\rm
tot}\propto g_{a\gamma'\gamma}^4$, this defines a thermal relic coupling
$g_{a\gamma'\gamma}^{\rm th}(m_a)$. At $m_{\gamma'}=300$~GeV, it falls
from $0.43$~GeV$^{-1}$ at $m_a=5$~GeV to $4.8\times10^{-3}$
~GeV$^{-1}$ at $100$~GeV and $1.2\times10^{-3}$~GeV$^{-1}$ at
$m_a\simeq300$~GeV 
as shown by the black dashed line in Fig.~\ref{fig:dm}. 
Larger couplings give a subdominant thermal relic.
Smaller couplings overproduce DM unless the cosmological history is
non-standard.

This estimate includes only $aa$ annihilation. As $m_a \simeq m_{\gamma'}$,
the dark photon remains in equilibrium 
through the portal vertex, via $\gamma'\leftrightarrow a\gamma$, and
coannihilation processes such as $a\gamma'\to f\bar f$ via
$s$-channel $\gamma/Z$, which scale as $g_{a\gamma'\gamma}^2$ rather
than $g_{a\gamma'\gamma}^4$, can dominate. The thermal target is
therefore reliable only for mass splittings
$(m_{\gamma'}-m_a)/m_a\gtrsim10\%$, i.e.\ $m_a\lesssim270$~GeV at
$m_{\gamma'}=300$~GeV.

\subsection{Collider vs.\ indirect detection}
\label{sec:dm-id}

To compare collider and indirect-detection sensitivities as a function of
the DM mass, we repeat the full collider analysis in an $m_a$ scan at
fixed $m_{\gamma'}=300$~GeV with the same simulation chain,
statistical treatment, and HL-LHC projection as in
Sec.~\ref{sec:results}. 
The collider limit is nearly flat for $m_a\lesssim100$~GeV and
weakens as $m_a$ approaches $m_{\gamma'}$. In the dark photon rest frame the
photon energy, $(m_{\gamma'}^2-m_a^2)/(2m_{\gamma'})$, shrinks and
pushes the lab-frame $E_T^\gamma$ spectrum of the on-shell
contribution below the SR threshold. Its SR efficiency falls from
$19.7\%$ at $m_a=1$~GeV to $9.1\%$ at $250$~GeV and $0.9\%$ at
$285$~GeV. Close to the endpoint, however, the far off-shell
contribution, $q\bar q\to a\gamma'^*$ with $\gamma'^*\to a\gamma$ at
virtuality well above $m_{\gamma'}$, takes over. Its photons are hard
(SR efficiency $\simeq35\%$), and its rate grows as
$g_{a\gamma'\gamma}^4$ rather than $g_{a\gamma'\gamma}^2$. It provides
$44\%$ of the SR signal at the LHC limit for $m_a=270$~GeV and
essentially all of it above $285$~GeV. The limit therefore levels off
at $g_{a\gamma'\gamma}^{(95)}\simeq4.8\times10^{-3}$~GeV$^{-1}$ instead
of collapsing, as it would in the narrow-width approximation.

\begin{figure}[t]
\centering
\includegraphics[width=\columnwidth]{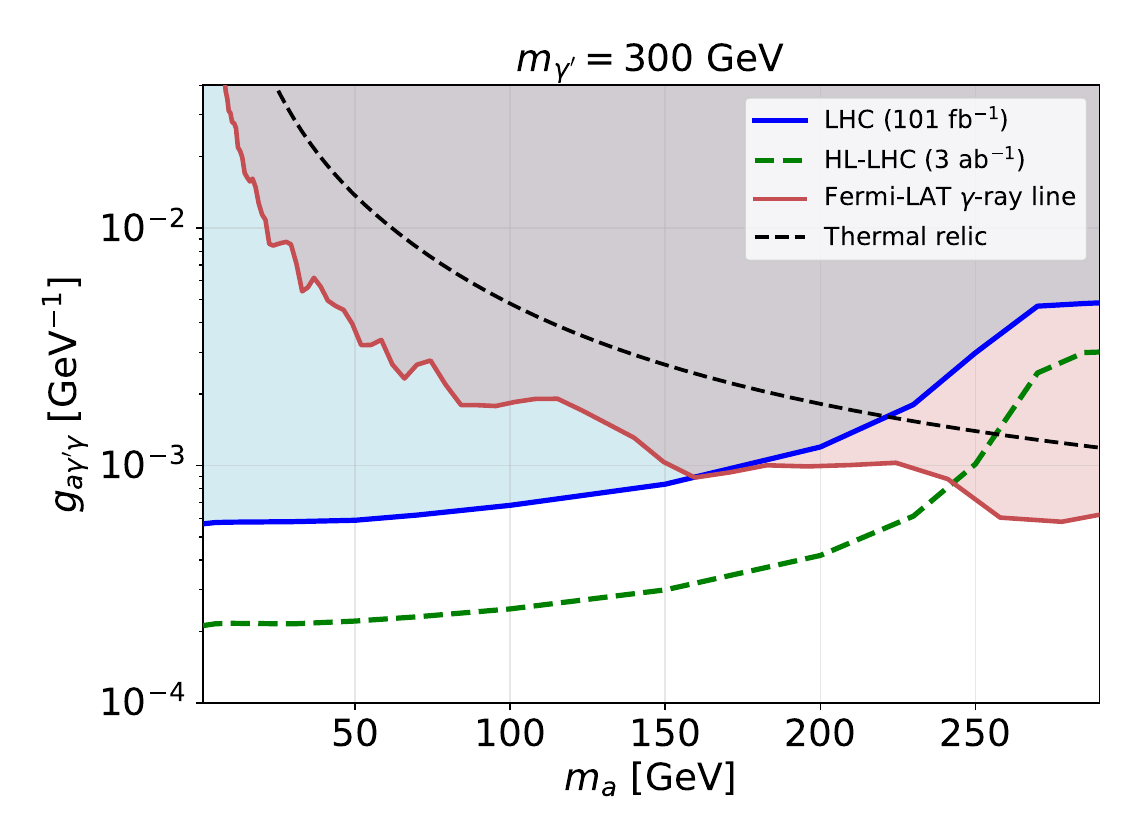}
\caption{Constraints in the $(m_a,g_{a\gamma'\gamma})$ plane at fixed
$m_{\gamma'}=300$~GeV: the LHC mono-photon exclusion (solid blue) and its
HL-LHC projection (dashed green); the Fermi-LAT gamma-ray line
constraint on $aa\to\gamma\gamma$ assuming $a$ constitutes all of the
DM (solid red; Fermi-LAT Pass~8~\cite{Fermi-LAT:2015kyq}); and the thermal relic coupling for
which $\langle\sigma v\rangle_{\rm tot}=3\times
10^{-26}$~cm$^3$\,s$^{-1}$ (black dashed). Shaded regions are
excluded.}
\label{fig:dm}
\end{figure}

Treating this region required care. Near the endpoint the dark photon
is extremely narrow at the limit couplings,
$\Gamma_{\gamma'}\simeq0.05$--$14$~MeV. MadGraph concentrates its
phase-space sampling around the Breit--Wigner peak. Here, however, the
on-shell photons are too soft to pass the pre-selection, and the
selected events come from the far off-shell region, which is then
sampled too sparsely. As a result the off-shell contribution is
underestimated by a factor of $\sim3$--$4$, as we verified by comparing
the hard-photon rate with the procedure described below. That region
does not depend on the width, so it can be computed reliably with a
larger fixed width. For $m_a\geq250$~GeV we therefore compute
the two contributions separately. Since every diagram contains a single
$\gamma'$ propagator, the pre-selected cross section can be written, up
to corrections of order $\Gamma_{\gamma'}/m_{\gamma'}$, as
$\sigma(g_{a\gamma'\gamma},\Gamma_{\gamma'})=g_{a\gamma'\gamma}^4\,[\,T+P/(m_{\gamma'}\Gamma_{\gamma'})\,]$ where we
denote $T$ the off-shell (tail) coefficient and $P$ the on-shell (peak)
coefficient. Both depend on $m_{\gamma'}$, $m_a$, and the selection,
but not on $g_{a\gamma'\gamma}$ or $\Gamma_{\gamma'}$. The first term is
the off-shell contribution, which does not depend on the width. The
second is the on-shell resonance, whose Breit--Wigner integral gives
$\pi/(m_{\gamma'}\Gamma_{\gamma'})$. With the physical width $\Gamma_{\gamma'}\propto g_{a\gamma'\gamma}^2$,
the on-shell term scales as $g_{a\gamma'\gamma}^2$ and the off-shell term as $g_{a\gamma'\gamma}^4$. The
off-shell part comes from the full process with a fixed width
$\Gamma_{\gamma'}=1$~GeV, after subtracting the peak contribution; the result is
stable to $1\%$ for fixed widths between $10$~MeV and $10$~GeV. The on-shell part comes from
the decay chain $pp\to a\gamma'$, $\gamma'\to a\gamma$ with the
physical width, for which the narrow-width approximation is exact. Each
part has its own simulated sample, and the limit is found by solving
the $CL_s$ condition directly as a function of $g_{a\gamma'\gamma}$.
At $m_a=250$~GeV, where the off-shell part is only $5\%$ of the SR
signal, this agrees with the standard treatment to $2\%$. The LHC
exclusion and HL-LHC sensitivity are shown by the solid blue and dashed
green lines in Fig.~\ref{fig:dm}, respectively.

For indirect detection, we use the Fermi-LAT gamma-ray line
search~\cite{Fermi-LAT:2015kyq}. It uses 5.8 years of Pass~8 data to
set 95\% CL upper limits on $\langle\sigma v\rangle_{\gamma\gamma}$
for line energies from $0.2$ to $\sim500$~GeV. We use its observed limit
for the NFW profile and the R41 region of interest. 
Assuming the axion constitutes all of the DM, a limit on the
coupling can be given by 
\begin{equation}
g_{a\gamma'\gamma}(m_a)\leq g_{a\gamma'\gamma}^{\rm ref}
\left(\frac{\langle\sigma v\rangle^{\rm Fermi}_{\gamma\gamma}(m_a)}
{\langle\sigma v\rangle_{\gamma\gamma}(m_a;\,g_{a\gamma'\gamma}^{\rm
ref})}\right)^{1/4},
\end{equation}
where $g_{a\gamma'\gamma}^{\rm ref}$ is an input reference coupling strength.

The Fermi-LAT constraint is shown by the red line in Fig.~\ref{fig:dm}. 
The collider and indirect detection sensitivities have opposite mass dependence. 
The collider limit weakens as $m_a$ approaches to $
m_{\gamma'} = 300$ GeV, while the indirect-detection limit strengthens steeply
with $m_a$, because $\langle\sigma v\rangle_{\gamma\gamma}\propto
m_a^6$ at $m_a\ll m_{\gamma'}$. 
Indirect detection
reaches the current LHC limit at $m_a\simeq160$~GeV, stays within
$5\%$ of it up to $m_a\simeq183$~GeV, and is clearly stronger above
$m_a\simeq190$~GeV. The indirect detection limit falls below the HL-LHC projection at
$m_a\simeq241$~GeV. At
$m_a=298.6$~GeV, Fermi-LAT constraints
$g_{a\gamma'\gamma} < 6.5\times10^{-4}$
~GeV$^{-1}$. That is about $7.5$ times stronger than the LHC limit
there ($4.9\times10^{-3}$~GeV$^{-1}$) and about $4.6$ times stronger
than the HL-LHC projection ($3.0\times10^{-3}$~GeV$^{-1}$).
The collider search thus dominates for light axions, while indirect
detection closes the nearly degenerate window $m_a\simeq m_{\gamma'}$
that the mono-photon search cannot reach.

The thermal relic line sharpens this picture. The LHC excludes the
thermal relic coupling for $m_a\lesssim221$~GeV, and the HL-LHC would
extend this to $m_a\lesssim256$~GeV. On the thermal line the axion
makes up exactly the observed DM, so the 100\%-DM assumption of the
indirect-detection limit holds by construction. There, the Fermi-LAT
limit lies below the thermal relic coupling by a factor between $0.17$ and $0.65$ in
$g_{a\gamma'\gamma}$. A thermal relic axion is therefore excluded 
across the entire axion mass range considered for $m_{\gamma'}=300$~GeV.
The weakest margin is at $m_a\simeq225$~GeV, where the predicted
$\langle\sigma v\rangle_{\gamma\gamma}$ exceeds the Fermi-LAT limit
by a factor of $\simeq6$. Near the endpoint, where
the coannihilation caveat of Sec.~\ref{sec:dm-thermal} applies, the
excess is $\simeq10$.

This exclusion does not depend on the value of $g_{a\gamma'\gamma}$ and
can be seen directly in terms of cross sections. The axion couples only
through the $a\gamma'\gamma$ and $a\gamma'Z$ vertices, so its only
tree-level annihilation channels are $\gamma\gamma$, $\gamma Z$, and
$ZZ$. There is no tree-level annihilation into SM fermions. The
monochromatic $\gamma\gamma$ channel therefore carries a large fraction of the total rate. From
Eqs.~\eqref{eq:sigmavgg}--\eqref{eq:sigmavzz}, $f_{\gamma\gamma}= \langle\sigma v\rangle_{\gamma\gamma}/\langle\sigma v\rangle_{\rm tot} = 1$ for
$m_a<m_Z/2$, and $f_{\gamma\gamma}\geq0.61$ for any $m_{\gamma'}>m_a$
throughout $m_a\leq300$~GeV, because the other channels are suppressed
by $t_W^2$ and $t_W^4$. A thermal relic thus predicts
$\langle\sigma v\rangle_{\gamma\gamma}=f_{\gamma\gamma}\times3\times
10^{-26}$~cm$^3$\,s$^{-1}$. Line searches, designed for the
loop-suppressed photon lines of conventional WIMPs, probe
$10^{-29}$--$10^{-27}$~cm$^3$\,s$^{-1}$ in this mass range. The
predicted rate exceeds the Fermi-LAT limit by a factor of $\sim130$--
$1200$ for $m_a\lesssim50$~GeV and $\sim6$--$60$ for
$m_a=100$--$300$~GeV. 
The exceptions are for axion mass belows the constraint data ($m_a < 0.2$ GeV for Fermi-LAT data)
or where coannihilation with a nearly degenerate $\gamma'$ controls
freeze-out or where the cosmological history is non-standard. 
A detailed study of these cases is left for future work. 

Away from the thermal-relic line, the Fermi-LAT exclusion region in Fig.~\ref{fig:dm} 
should be interpreted with care, because it assumes $\Omega_a=\Omega_{\rm DM}$
at every coupling. If the axion is a thermal relic, couplings below the thermal relic value (denoted as $g_{a\gamma'\gamma}^{\rm th}$, the black dashed curve in Fig.~\ref{fig:dm}) lead to DM overproduction and are therefore excluded in standard cosmology. Couplings above  the thermal relic value give a subdominant relic with
$\Omega_a\propto g_{a\gamma'\gamma}^{-4}$, so the line signal,
$\propto\Omega_a^2\langle\sigma v\rangle$, falls as
$(g_{a\gamma'\gamma}^{\rm th}/g_{a\gamma'\gamma})^4$. For a thermal
axion, Fermi-LAT therefore excludes only a band above the thermal
line, extending to about $5.9\,g_{a\gamma'\gamma}^{\rm th}$ at
$m_a=5$~GeV and about $1.8\,g_{a\gamma'\gamma}^{\rm th}$ at
$m_a=300$~GeV. Larger couplings are constrained only by the collider
searches.

\section{Conclusion}
\label{sec:conclusion}

We have constrained the
axion--photon--dark photon interaction using the CMS mono-photon data. 
We simulate the full process $pp\to aa\gamma$ with the finite dark
photon width, which is not captured by the narrow-width approximation for sufficiently large
$m_{\gamma'}$. The 95\% CL exclusion on $g_{a\gamma'\gamma}$ reaches
$4.8\times10^{-4}$~GeV$^{-1}$ for $m_{\gamma'}\lesssim100$~GeV and
extends up to $m_{\gamma'}\simeq2.3$~TeV, where the required coupling
reaches the limit of validity of the effective theory. For heavier dark photons the
width at the relevant couplings exceeds the mass, the signal rate
saturates, and no coupling is excluded.
This improves upon the LEP mono-photon bound by a
factor of $3$ or more across that bound's entire range. The
HL-LHC at $14$~TeV will improve it by a factor of $\sim2.7$ and extend
it to $m_{\gamma'}\simeq3$~TeV. A polarized
ILC250 would be more sensitive at light dark photon masses, but only a
hadron collider probes $m_{\gamma'}\gtrsim250$~GeV.

If the axion is the dark matter, the same vertex induces $s$-wave
annihilation to $\gamma\gamma$, $\gamma Z$, and $ZZ$. 
We found that the collider and the Fermi-LAT gamma-ray line
search are complementary. For $m_{\gamma'}=300$~GeV, 
the LHC is stronger for $m_a\lesssim160$~GeV,
and the line search is up to about $7.5$ times stronger near the
kinematic endpoint $m_a\simeq m_{\gamma'}$, where the collider
sensitivity is carried by the off-shell dark photon. 
Together, these constraints exclude a thermal-relic axion 
across the mass range considered at this benchmark.

For heavier dark photons, the endpoint window accessible to the collider search shifts to larger axion masses. 
In this regime, indirect detection limits from H.E.S.S.~\cite{HESS:2026ila}, CTAO~\cite{CTAO:2024wvb},  
MAGIC~\cite{MAGIC:2022acl} can be applied. 
A joint analysis of the $\gamma\gamma$ and $\gamma Z$ lines, as demonstrated using 15 years of Fermi-LAT data~\cite{DeLaTorreLuque:2023fyg}, could test their predicted ratio. Near $m_a\simeq300$~GeV the two lines are separated by only $\Delta E/E=m_Z^2/(4m_a^2)\simeq2\%$, which requires the percent-level energy resolution of proposed instruments such as HERD~\cite{Huang:2015fca}.

\begin{acknowledgments}
This work was partially supported by the National Science and Technology Council (NSTC) of Taiwan under Grant No. NSTC-115-2112-M-003-014 (CRC), the Ministry of Education (Higher Education Sprout Project NTU-114L104022-1), and the National Center for Theoretical Sciences of Taiwan (VQT).
This work was also supported in part by the Vietnam National Foundation for Science and Technology Development (NAFOSTED) under grant number 103.01-2023.50 (VQT).
\end{acknowledgments}

\appendix

\section{Additional validation material}
\label{app:validation}

We validate the simulation chain of Sec.~\ref{sec:method} against two
signal benchmarks for which CMS publishes exact selection efficiencies
(acceptance$\times$efficiency) in its HEPData record~\cite{CMS:2025cgw}.
Each benchmark is implemented as a UFO model, generated at LO with
MadGraph5~\cite{Alwall:2014hca}, showered with Pythia8~\cite{Sjostrand:2014zea},
passed through Delphes~\cite{deFavereau:2013fsa}, and analyzed with the
same SR selection and $80\%$ photon identification scale factor used for our signal.
We generate $4\times10^4$ events per point.
\begin{itemize}
\item \emph{EW-DM contact interaction}: $q\bar q\to Z/\gamma^*\to
  \gamma\chi\bar\chi$ through the dimension-seven
  operator of Ref.~\cite{Nelson:2013pqa}, with suppression scale
  $\Lambda=1$~TeV, at $m_\chi=1$ and $400$~GeV.
\item \emph{Simplified DM with a vector mediator}: a $Z'$ with universal
  vector couplings $g_q=0.25$ to quarks and $g_{\rm DM}=1$ to a Dirac
  fermion $\chi$, with $M_{Z'}=1$~TeV, at $m_\chi=1$ and $150$~GeV.
\end{itemize}

For the simplified-DM benchmark, CMS publishes the cutflow beginning with the
$E_T^{\rm miss}>200$~GeV requirement~\cite{CMS:2025cgw}; no
cutflow is published for the EW-DM benchmark. Table~\ref{tab:cutflowval}
compares the cumulative survival fractions, normalized to the
$E_T^{\rm miss}>200$~GeV step. Every
step agrees within about $0.5$ percentage points. The largest differences
occur at the lepton veto, where our Delphes-level lepton identification
is simpler than CMS's loose identification, and at the final
$E_T^\gamma/E_T^{\rm miss}$ requirement.

\begin{table}[h]
\centering
\caption{Cumulative survival fractions (\%) for the vector
simplified-DM benchmark ($M_{Z'}=1$~TeV), normalized to the
$E_T^{\rm miss}>200$~GeV step: CMS~\cite{CMS:2025cgw} vs.\ this work.}
\label{tab:cutflowval}
\small
\begin{tabular}{lcccc}
\toprule
 & \multicolumn{2}{c}{$m_\chi=1$~GeV} & \multicolumn{2}{c}{$m_\chi=150$~GeV} \\
Cut & CMS & Ours & CMS & Ours \\
\midrule
$E_T^{\rm miss}>200$~GeV & 100 & 100 & 100 & 100 \\
$\Delta\phi(\gamma,E_T^{\rm miss})>0.5$ & 99.71 & 99.74 & 99.68 & 99.73 \\
Lepton veto & 99.26 & 99.71 & 99.27 & 99.64 \\
$\Delta\phi(j,E_T^{\rm miss})>0.5$ & 90.53 & 90.43 & 90.28 & 90.28 \\
$E_T^\gamma/E_T^{\rm miss}<1.4$ & 89.05 & 88.77 & 89.05 & 88.53 \\
\bottomrule
\end{tabular}
\end{table}

Table~\ref{tab:validation} compares our overall efficiencies with those of CMS.
Before the photon-ID scale factor, our efficiencies are higher than CMS's
by $20$--$35\%$, because Delphes does not simulate the
$\sigma_{i\eta i\eta}$ shower-shape requirement. With the scale factor
applied, the agreement is within $8.3\%$ for the EW-DM benchmark and
within $4.2\%$ for the simplified-DM benchmark. The agreement improves
with DM mass in both cases.

\begin{table}[h]
\centering
\caption{Signal-region efficiency of the CMS benchmark models in our
simulation chain, before (raw) and after the $80\%$ photon-ID scale
factor (SF), compared with the CMS values from HEPData~\cite{CMS:2025cgw}.
Efficiencies are in \%, $m_\chi$ in GeV. The last column is the
relative difference between our SF-corrected efficiency and CMS's.}
\label{tab:validation}
\small
\begin{tabular}{lccccc}
\toprule
Model & $m_\chi$ & Raw & SF & CMS & Diff. \\
\midrule
EW-DM     & 1   & 49.4 & 39.5 & 36.45 & $+8.3\%$ \\
EW-DM     & 400 & 54.1 & 43.3 & 42.01 & $+3.1\%$ \\
Simp.\ DM & 1   & 20.4 & 16.3 & 17.05 & $-4.2\%$ \\
Simp.\ DM & 150 & 20.1 & 16.1 & 16.20 & $-0.5\%$ \\
\bottomrule
\end{tabular}
\end{table}

\allowdisplaybreaks
\bibliographystyle{apsrev4-2}
\bibliography{refs}

\end{document}